\documentclass[%
 twocolumn,
*preprint,
preprintnumbers,
 amsmath,amssymb,
 aps,
]{revtex4-1}

\usepackage{graphicx}
\usepackage{epstopdf}
\usepackage{dcolumn}
\usepackage{bm}

\usepackage{color}

\begin{document}

\title{Spontaneous mirror-symmetry breaking in a photonic molecule}

\author{Philippe Hamel}
\affiliation{Laboratoire de Photonique et de Nanostructures (CNRS
UPR 20),Route de Nozay, Marcoussis, 91460, France }

\author{Samir Haddadi}
\affiliation{Laboratoire de Photonique et de Nanostructures (CNRS
UPR 20),Route de Nozay, Marcoussis, 91460, France }

\author{Fabrice Raineri}
\affiliation{Laboratoire de Photonique et de Nanostructures (CNRS
UPR 20),Route de Nozay, Marcoussis, 91460, France }

\author{Paul Monnier}
\affiliation{Laboratoire de Photonique et de Nanostructures (CNRS
UPR 20),Route de Nozay, Marcoussis, 91460, France }

\author{Gregoire Beaudoin}
\affiliation{Laboratoire de Photonique et de Nanostructures (CNRS
UPR 20),Route de Nozay, Marcoussis, 91460, France }

\author{Isabelle Sagnes}
\affiliation{Laboratoire de Photonique et de Nanostructures (CNRS
UPR 20),Route de Nozay, Marcoussis, 91460, France }

\author{Ariel Levenson}
\affiliation{Laboratoire de Photonique et de Nanostructures (CNRS
UPR 20),Route de Nozay, Marcoussis, 91460, France }

\author{Alejandro M. Yacomotti}
\email{Alejandro.Giacomotti@lpn.cnrs.fr}
\affiliation{Laboratoire de
Photonique et de Nanostructures (CNRS UPR 20),Route de Nozay,
Marcoussis, 91460, France }

\date{\today}
\begin{abstract}

Multi-cavity photonic systems, known as photonic molecules (PMs), are ideal multi-well potential building blocks for advanced quantum and nonlinear optics\cite{Abbarchi:2013fk,Dousse:2010ys,Gerace:2009vn,Liew:2010uq}. A key phenomenon arising in double well potentials is the spontaneous breaking of the inversion symmetry, i.e. a transition from a delocalized to two localized states in the wells, which are mirror images of each other. Although few theoretical studies have addressed mirror-symmetry breaking in micro and nanophotonic systems \cite{rodrigues2013symmetry, Maes:05, Maes:06, Bulgakov:12, bulgakov2013light}, no experimental evidence has been reported to date. Thanks to the potential barrier engineering implemented here, we demonstrate spontaneous mirror-symmetry breaking through a pitchfork bifurcation in a PM composed of two coupled photonic crystal nanolasers.
Coexistence of localized states is shown by switching them with short pulses. This offers exciting prospects for the realization of ultra-compact, integrated, scalable optical flip-flops based on spontaneous symmetry breaking. Furthermore, we predict such transitions with few intracavity photons for future devices with strong quantum correlations.

\end{abstract}


\maketitle


Spontaneous symmetry breaking (SSB) unifies diverse physical mechanisms through which a given symmetric system ends up in an asymmetric state\cite{golubitsky1988singularities}.
It explains many central questions from particle and atomic physics to nonlinear optics (the Goldstone boson and the Higgs mechanism\cite{strocchi2005symmetry,Endres:2012nx}, phase transitions in Bose-Einstein condensates --BECs--\cite{sadler2006spontaneous,Zibold:2010kl}, metamaterials\cite{Liu:2014kx}, bifurcations in lasers\cite{Green:1990kx,Ackemann:2013ys}, photorrefractive media\cite{Kevrekidis:2005bh}, to mention just a few).
A paradigmatic symmetry in this context is given by reflection in a double-well potential (DWP), as it is the case of pyramidal molecules (e.g. ammonia)\cite{jona2002interaction}: SSB dictates whether the state of a system will be delocalized or, in turn, confined within either well.
In photonics, such a mechanism is possible provided the third order nonlinearities overcome photon tunneling\cite{Malomed:2013fk}.
In this work we experimentally show SSB in a photonic molecule (PM) given by two evanescently coupled photonic crystal (PhC) nanolasers. Switchable localized modes with broken mirror-symmetry will be demonstrated herein. This can be prospected as a nanoscale version of a laser flip-flop\cite{Liu:2010fk}; the memory is pumped incoherently, set and reset can be induced with positive pulses and there is no coherent driving beam to bias the device, as in conventional bistable cavities. This paves the way for the realization of ultra-small flip-flop optical memories based on SSB.

\begin{figure}[!h]
\centering
\includegraphics[scale=0.3,clip=true]{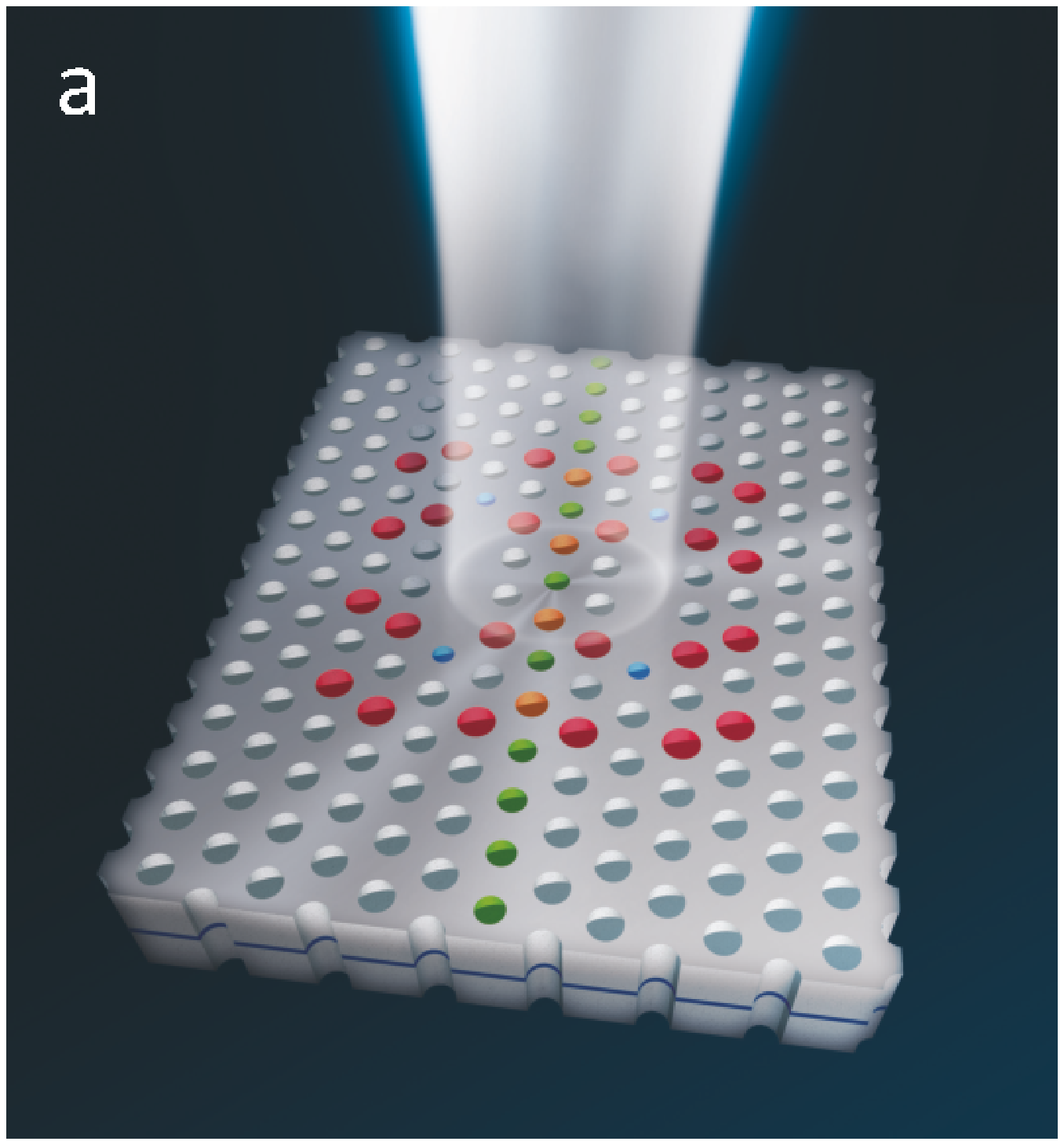}
\includegraphics[scale=0.3,clip=true]{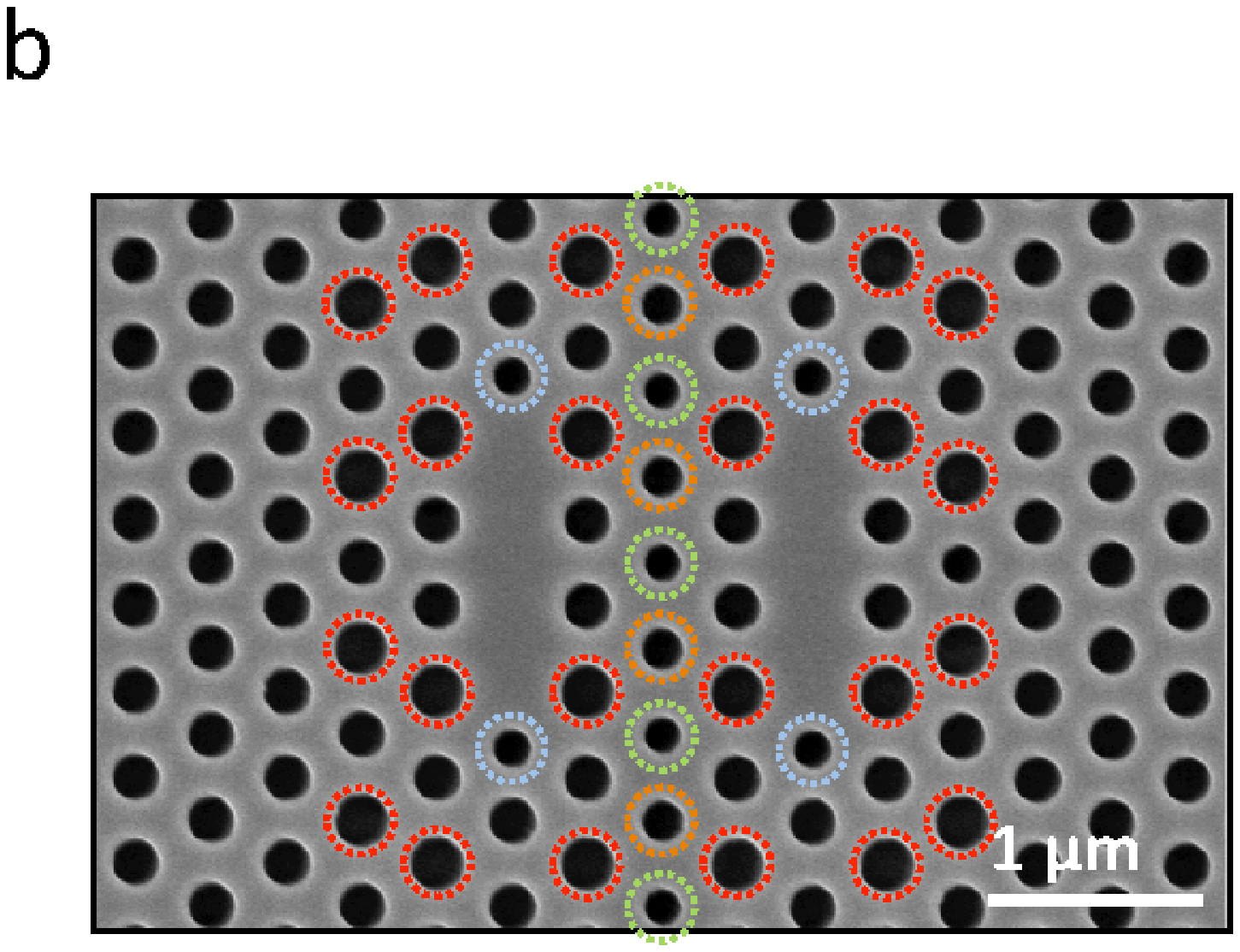}
\includegraphics[scale=0.35,clip=true]{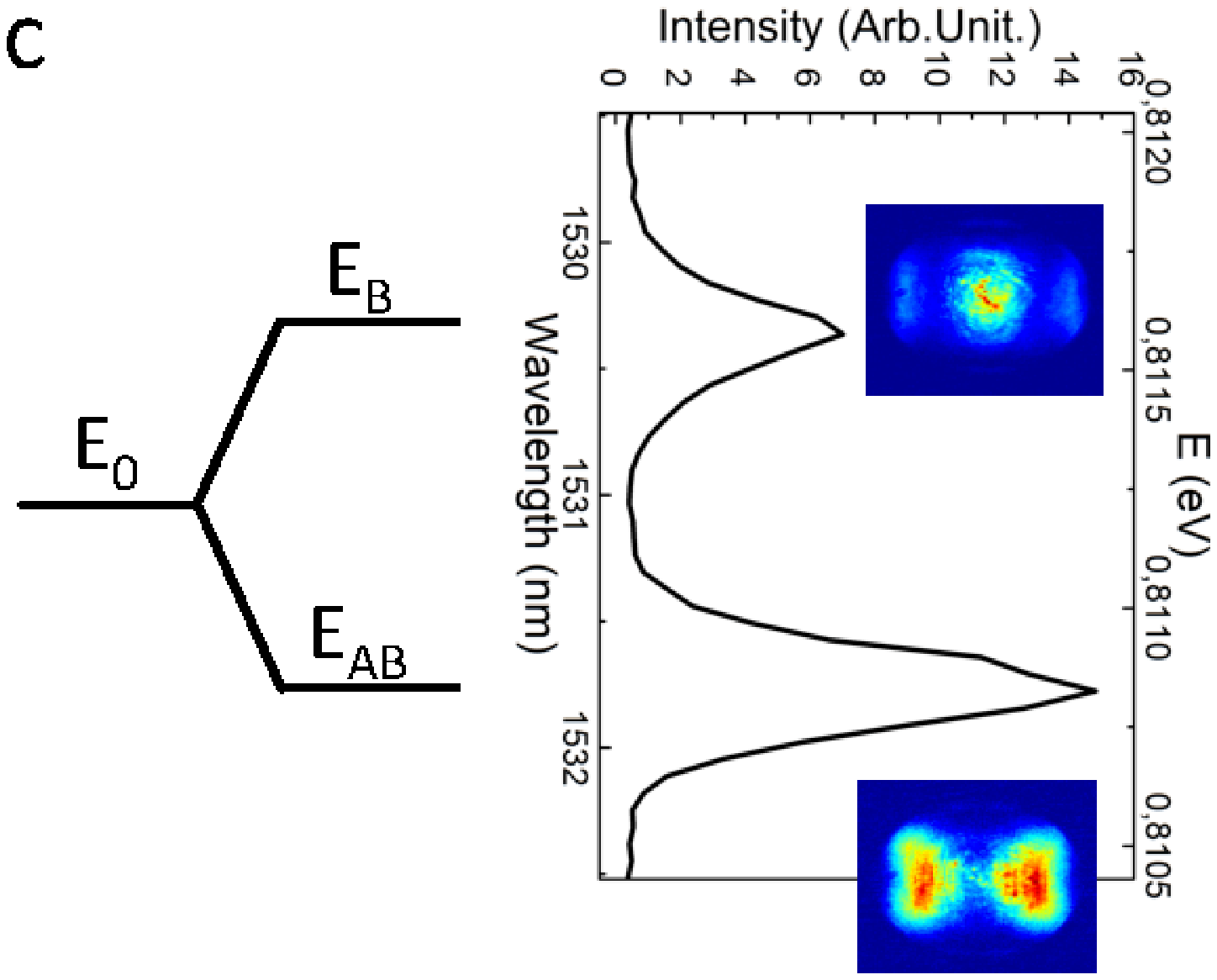}
\caption{{\bf Photonic molecule}. {\bf a}, 3D sketch of the photonic molecule composed of two coupled L3 PhC nanocavites. PhC lattice period is $a=425$ nm, hole radius $r=0.266a$.  Embedded QWs are represented by a dark blue line. Blue holes ($r_{blue}=r-0.06a$), shifted in the $\Gamma$K direction by $\Delta r_{blue}=0.15a$, increase the cavity Q-factor. Red holes ($r_{red}=r+0.05a$)  improve beaming of the radiated photons. Green holes ($r_{green}=r-20\%$) control the coupling strength. Orange holes ($r_{orange}=r_{red}-20\%$) combine both effects. {\bf b}, SEM image of the fabricated sample; dashed circles highlight the modified holes. {\bf c}, Splitted energy levels of the PhC molecule (note that the ground state is the anti-bonding mode). Insets: far-field emission profiles of bonding (B, top) and anti bonding (AB, bottom) modes.} \label{coupled_cavities}
\end{figure}

We represent the PM as a DWP, symmetric with respect to the inversion plane. We describe the dynamics in terms of the complex amplitudes of the photonic field at the left ($\psi_L$) and right ($\psi_R$) sites, $|\psi|^2$ being photon number. A finite potential barrier leads to a tunneling rate $K$. We further consider a local (nonlinear) interaction $U |\psi_{L,R}|^2$, and a lifetime $\tau$ due to losses. SSB instabilities occur as long as $K$ is lower than a critical value $K_c$ ($|K|<|K_c|$), with $|K_c\tau|\sim |U| \cdot |\psi|^2$\cite{PhysRevE.64.025202}. In the case of our PM laser, $|\psi|^2$ will linearly increase with the pump power $P_p$ after the threshold power $P_{th}^-$ corresponding to the anti-bonding mode, which is the hybrid mode of the PM minimizing the optical losses. Hence $|\psi|^2 \rightarrow |\psi_-|^2=\Delta P_-=P_p-P_{th}^-$. In a semiconductor medium, $U$ can be related to the phase-amplitude coupling factor $\alpha$ (the Henri factor); $K_c$ then yields $K_c \tau \sim \alpha \Delta P_-$, which can also be recast  as
\begin{equation}
\Delta P_c\sim K\tau/\alpha.
\label{critical}
\end{equation}
In absence of cavity detuning, $K$ is related to mode splitting ($\Delta \lambda _{split}$) as $K\tau=\Delta \lambda _{split}/\delta \lambda$, where $\delta \lambda$ is the cavity linewidth. On the other hand, $\alpha$ lies between $\sim 5$ and $10$. Hence a good DWP-candidate to demonstrate SSB with low pump powers has to fulfill the condition that mode splitting be of the order of the resonance width, i.e. $K\tau \sim1$. Finely controlling coupling strength in PMs is thus a key ingredient to achieve SBB transitions. We implement such a control --together with both efficient laser emission and free-space photon collection-- by means of an original PhC cavity design, as explained hereafter.

\begin{figure}[!h]
\centering
\includegraphics[scale=0.5,clip=true]{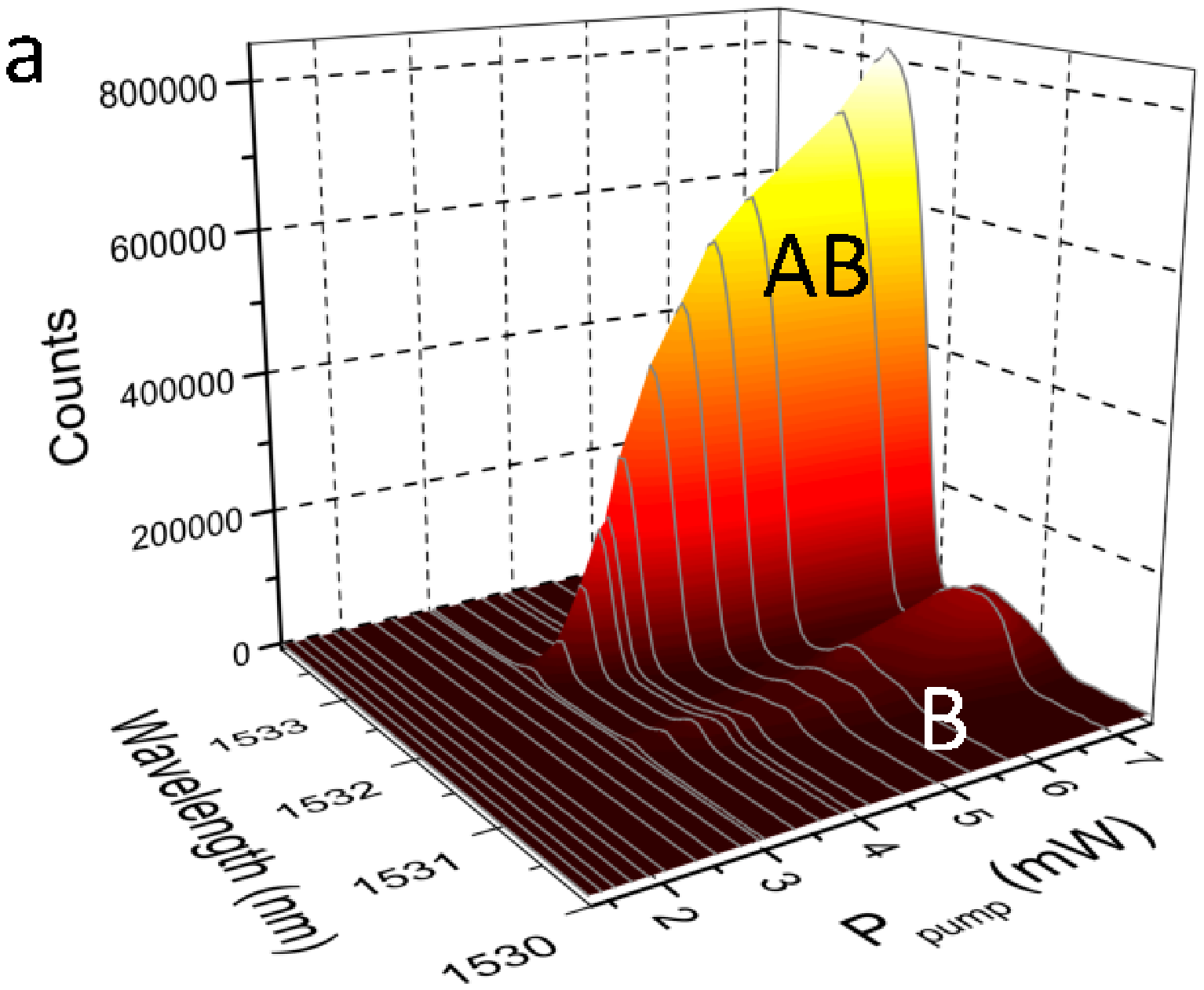}
\includegraphics[scale=0.5,clip=true]{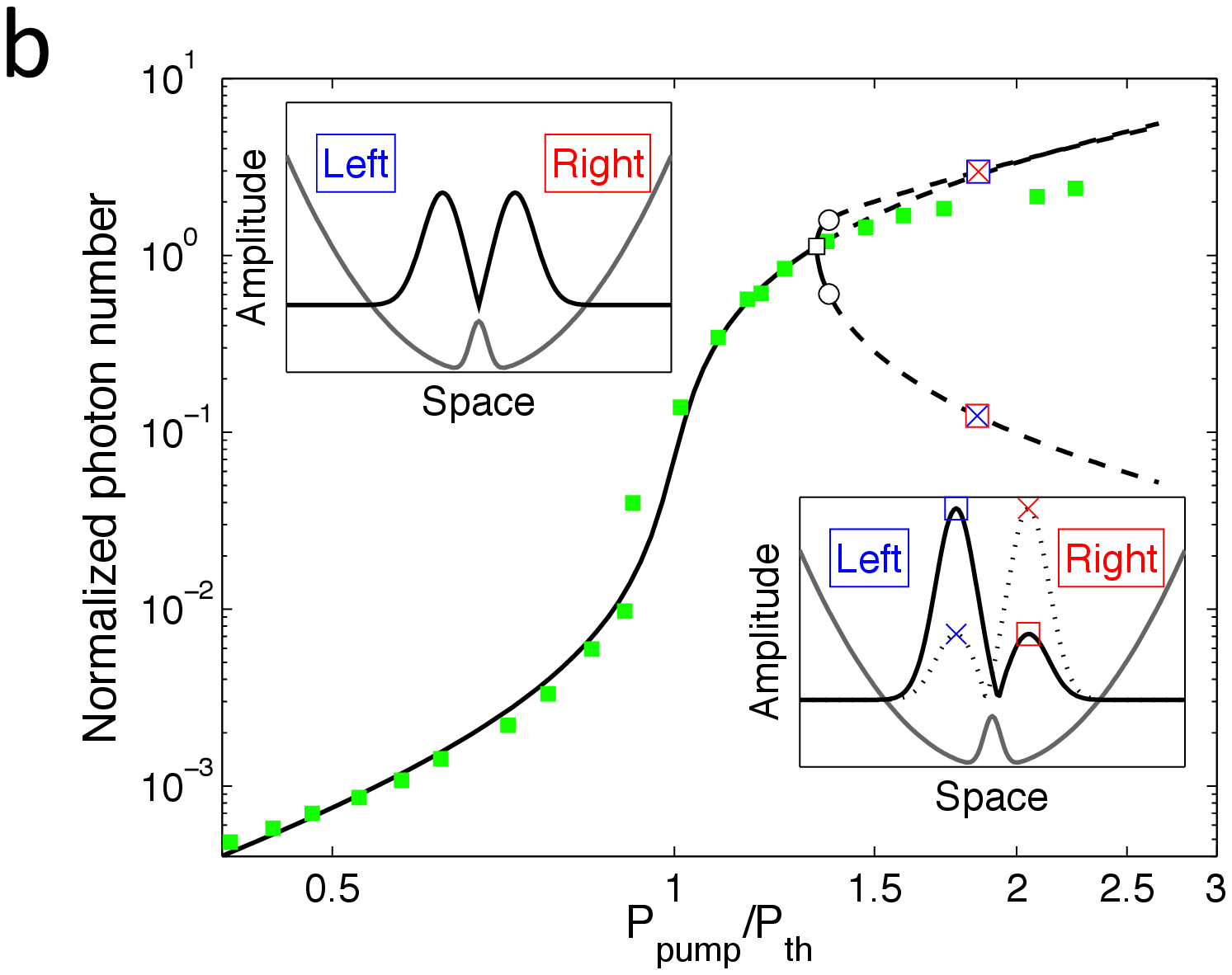}
\caption{{\bf Laser emission of coupled nanocavities}. {\bf a}, Spectral measurements as a function of pump power when pumping at the center of the system; both modes (B, short wavelength, and AB, higher wavelength) are observed, but only AB undergoes laser emission. {\bf b}, AB laser mode output vs input power. Green filled squares: experimental measurements obtained from spectral peaks in ({\bf a}) for increasing input power. Black line: Numerical solution of coupled lasers rate equations. Black open square: pitchfork bifurcation. Colored open squares: first broken parity state. Colored crosses: second broken parity state. Open circles: bifurcations leading to oscillations. Insets: illustrations of the double-well potential with the unique stable solution before bifurcation (left inset) and the two co-existing solutions after bifurcation (right inset). } \label{pitchfork}
\end{figure}

Our DWP landscape is obtained with the specific PhC molecule depicted in Fig.  \ref{coupled_cavities}a. This is formed by two evanescently coupled PhC L3 cavities (three holes missing in the $\Gamma$K direction of a triangular lattice) in a semiconductor free standing membrane (Fig. \ref{coupled_cavities}a,b). With the aim of realizing high-Q cavities with improved beaming and controlled coupling strength, three conception tools are used: i) end-holes of each L3-cavity are shifted and shrunk in order to increase theoretical Q-factors up to $\sim \, 10^5$ (blue holes in Fig. \ref{coupled_cavities}a,b)\cite{Akahane:2003zr,Portalupi:2010}; ii) the radius of neighbor holes are modified in order to confine radiated photons within a $\sim \, 30^\circ$ emission cone \cite{Haddadi2012,Haddadi2013} (red holes in Fig. \ref{coupled_cavities}a,b); iii) the hole size of the central row (green holes in Fig. \ref{coupled_cavities}a,b) is modified in order to control the coupling strength towards $K\tau\sim1$\cite{Haddadi:14}. Both single and coupled cavities have been etched on InP membranes containing InGaAs/InGaAsP quantum wells (see Methods). Resonant wavelengths are about $\lambda\sim 1540 \,\mathrm{nm}$, and measured Q-factors of bare cavities (i.e. at QW transparency) are $Q=4970$ ($\tau \sim 8 \, \mathrm{ps}$) for single, and $Q=4300$  ($\tau\sim 7 \, \mathrm{ps}$) for coupled cavities.

Mode splitting has been measured through room-temperature micro-photoluminescence spectroscopy. A cw pump beam
is focussed down to a $\sim 1.5 \, \mu \mathrm{m}$-diameter spot at the center of the coupled cavity system (see Methods). Two modes of the PM can be observed: the anti-bonding, "AB" (ground state), and bonding, "B" (excited state)\cite{ground}. Far-field patterns showing intensity maxima (B) and minima (AB) at $k=0$ are shown in Fig.  \ref{coupled_cavities}c. From a mode splitting of $\Delta \lambda_{split}\approx1.4\, \mathrm{nm}$ (Fig. \ref{coupled_cavities}c),
the normalized coupling constant is $K \tau=3.3$ .

A solitary nano-cavity laser exhibits a S-shaped, output vs. input power curve, where its sharpness is related to the spontaneous emission $\beta$ factor. Now, what is the expected behavior for two evanescently coupled nanolasers when pumped at the center of the PM? Out of the two hybrid modes, the lasing mode is the one with lower optical losses, i.e. the AB mode; the B mode is strongly attenuated (Fig. \ref{pitchfork}a). In Fig. \ref{pitchfork}b we depict the AB maxima (green symbols) superimposed to a numerical solution of coupled lasers rate equations, with $\beta=0.017$ given by a fit of the experimental points (black line). A S-shaped curve for the AB mode is observed up to $P_p=1.33 \, P_{th}$. Within this range the solution is delocalized in the DWP (Fig. \ref{pitchfork}b, left inset). Above this value two branches of steady state solutions come up (plus a third one being the destabilized AB mode), corresponding to two co-existing solutions: the "Left cavity on" together with the "Right cavity off" ("L1R0" from now on), and the "Left cavity off" together with the "Right cavity on" solution ("L0R1"), see Fig. \ref{pitchfork}b, right inset.
Unlike the AB mode, these new solutions have no defined parity: a SSB instability takes place at $P_p=1.33 P_{th}$ in the form of a pitchfork bifurcation (Fig. \ref{pitchfork}, black square). The two new branches (upper and lower) remain stable up to $P_{p}=1.37 \, P_{th}$ where the system undergoes secondary instabilities (Hopf bifurcations, Fig. \ref{pitchfork}, circles) leading to ultrafast oscillations (predicted frequencies $\sim \, 150-180 \, \mathrm{GHz}$ depending on the pump power), larger but close to the beating note $\nu_{beat}=K/\pi\sim \, 148 \, \mathrm{GHz}$. These can be related to ac Josephson oscillations\cite{Abbarchi:2013fk}. For a lasing AB mode in presence of self-focussing nonlinearities (positive nonlinear refractive index above QW transparency), such SSB scenario is only possible for a specific sign of the optical coupling parameter $K$ --positive with our sign convention-- which imposes a lower energy for the AB mode. The L3 cavity-based PM implemented here fulfills this requirement.

\begin{figure}[!h]
\centering
\includegraphics[scale=0.43,clip=true]{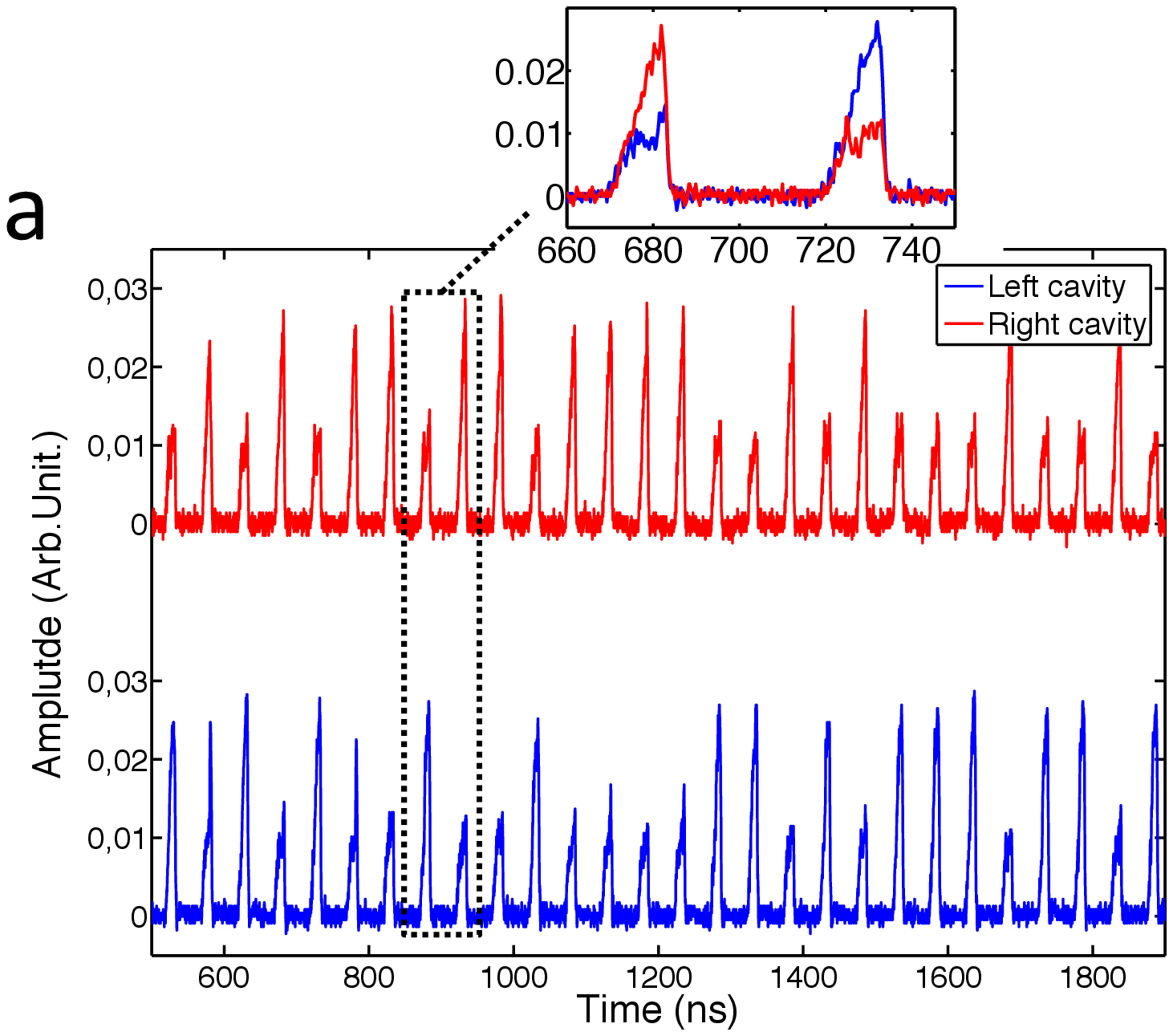}
\includegraphics[scale=0.43,clip=true]{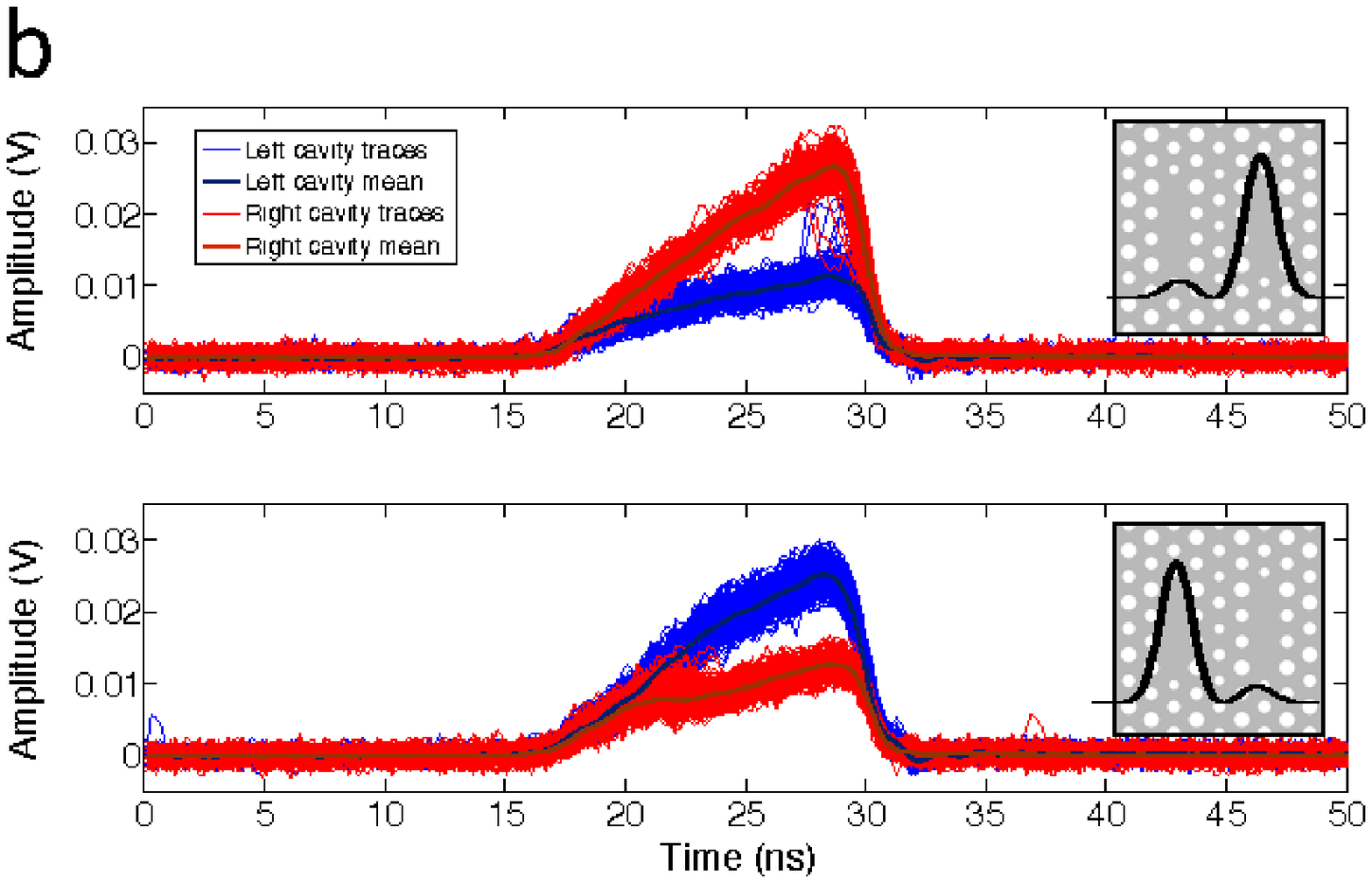}
\includegraphics[scale=0.43,clip=true]{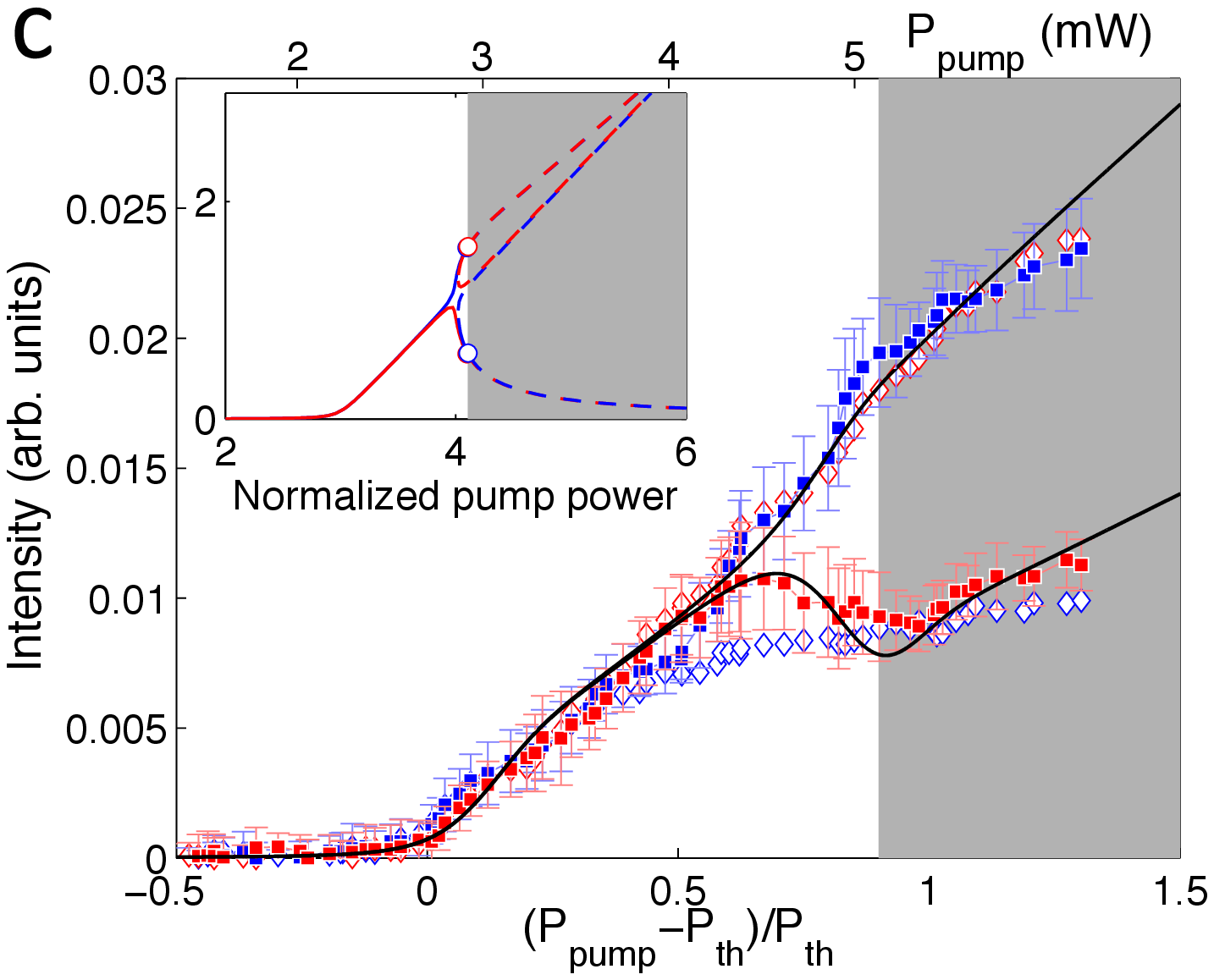}
\caption{{\bf Time domain measurements and pitchfork bifurcation}. {\bf a}, Pulse sequence of a $600\, \mu$s-duration output signal from both nanocavities (blue: L, red: R). The time series shows alternation of "High red-Low blue" and "Low red-High blue" states (dead time windows between pulses are omitted in a sequence). The zoom shows this alternation over two periods. (b) Superimposed time traces after identification of two different states, L1R0 and L0R1; averages are shown in thick line. (c) Same data as (b), plotted as a function of the instantaneous pump power, showing the SSB bifurcation (full squares: L1R0, open diamonds: L0R1). The black line shows a numerical (Runge-Kutta) integration of the coupled laser rate equations (x-axis is rescaled $\times2$ for a qualitative comparison with the experiment). Grey areas correspond to fast (filtered by the photodetector) oscillations in the time domain. The inset presents a theoretical bifurcation diagram with imperfect symmetry when the pump position is slightly shifted from the center of the molecule by $\delta p=2\times10^{-4}$ (pump beam slightly shifted to the R cavity).} \label{time_series}
\end{figure}

The spectral measurements presented in Fig. \ref{pitchfork} show a saturation of the integrated output laser power for $P_p>1.4 \, P_{th}$. This is consistent with the predicted bifurcation, since mean power for broken symmetry states is lower compared to the AB mode. In order to investigate SSB experimentally, fast time detection with spatial filtering of individual cavity outputs has been set up. Two identical APD photodiodes coupled to single-mode optical fibers are used to collect L and R cavity signals simultaneously. The diffraction limited collection area is smaller than the inter-cavity distance ($d=1.47 \, \mu \mathrm{m}$) such that less than 10\% cross-talk is observed (see Methods). The modulated pump beam (50 KHz, 30 ns-rise time), impinging the sample with a peak power of $\sim 6 \, \mathrm{mW}$, is aligned at the center of the PM.

Fig. \ref{time_series}a shows a sequence of simultaneous outputs from both cavities. Segments of alternating "High blue-Low red", and "Low blue-High red" peaks can be observed. We average out these time traces by superimposing events using a peak detection algorithm (see Methods). The result is shown in Fig. \ref{time_series}b. Two types of events are clearly identified: L1R0 and L0R1. These are plotted in Fig. \ref{time_series}b as function of time, and in Fig. \ref{time_series}c as a function of the instantaneous pump power, together with a numerical integration of the coupled lasers rate equations. The AB mode builds up from noise, with a laser threshold of ${\cal P}_{th}\approx 2.7 \, \mathrm{mW}$, and evolves in the usual way up to $({\cal P}_p-{\cal P}_{th})/{\cal P}_{th}\sim 0.7$ (${\cal P}_p\approx 4.5 \, \mathrm{mW}$) where the two distinct branches of output states come up. It is important to point out that the lower power branch, instead of monotonically decreasing as ${\cal P}_p$ is increased, raises again for $({\cal P}_p-{\cal P}_{th})/{\cal P}_{th}>1$. This is in good agreement with the the model, being a consequence of the fast oscillations.
Experimentally, $\sim100$ GHz oscillations are filtered out by the APD bandwidth and only the DC intensity component is measured, which is higher than the steady state intensity of the lower branch (Fig. \ref{time_series}c, inset).

In absence of external noise terms, a small shift of the pump beam from the center of the PM ($\delta p=2\times10^{-4}$ in the model) may be responsible for triggering one state or another within each pump pulse\cite{detuning} .
Experimentally, the fact that fast alternation between stable states is observed may be due to i) mechanical vibrations; and ii) spontaneous emission fluctuation noise making the system to spontaneously choose L1R0 or L0R1 states. In either case, a way to experimentally prove that our scenario indeed corresponds to SSB, i.e. that  L1R0 and L0R1 states {\em do coexist}, is to be able to switch from one to the other within the same pump pulse, as it will be done in the following.

\begin{figure}[ht!]
\centering
\includegraphics[scale=0.5]{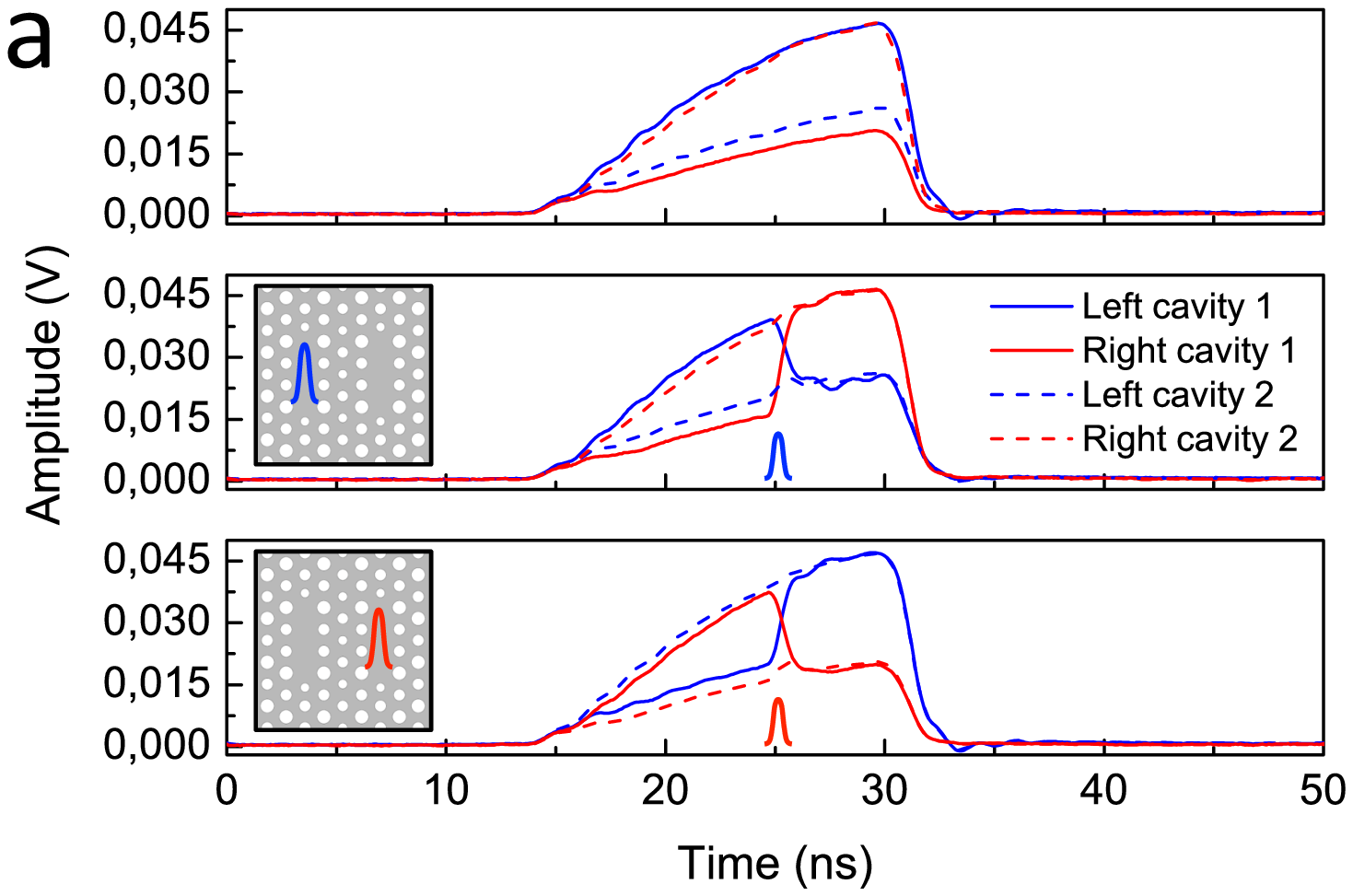}
\includegraphics[scale=0.5]{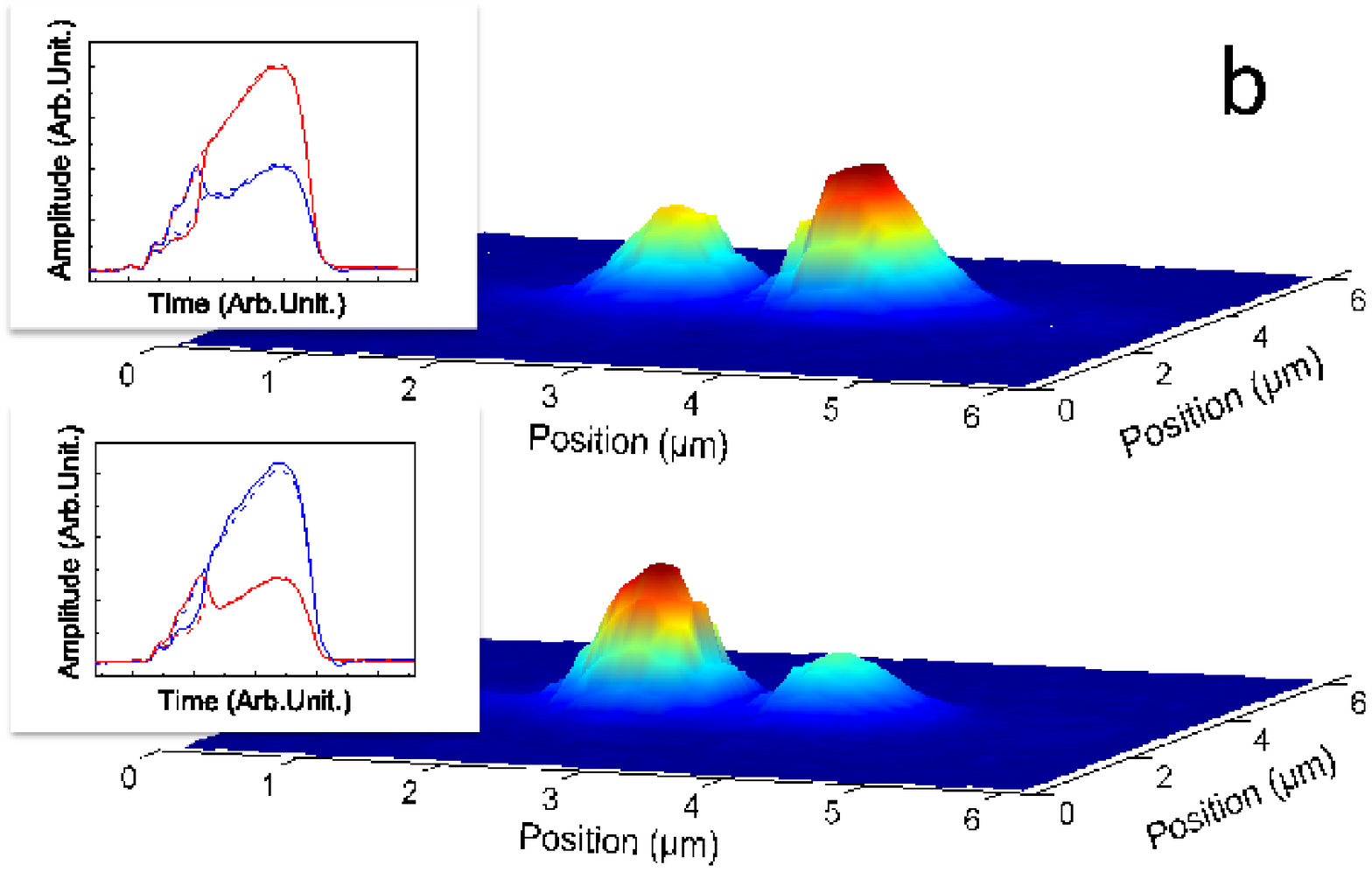}
\caption{{\bf Coexistence of parity broken states}. {\bf a}, Switching from one state to the other is triggered by a short ($100$ ps) pulse on each cavity. Top: non-perturbed spontaneous switching. Center: the pulse is applied to the L cavity (sketched as a blue peak). Bottom: the perturbation is applied to the R cavity (sketched as a red peak). Continuous and dashed lines: states "1" and "2" respectively, defined as the states before the arrival of the pulse ("1": L1R0 in the top and center frames, L0R1 in the bottom frame; "2": L0R1 in the top and center frames, L1R0 in the bottom frame). {\bf b}, Image of the intensity profile from an InGaAs camera after stabilizing states using an early perturbation. Insets: Time traces corresponding to early induced L0R1 (top) and L1R0 (bottom) states.} \label{switch_exp}
\end{figure}

Coexistence of L1R0 and L0R1 is experimentally investigated by means of an additional short ($\sim100 \, \mathrm{ps}$) pulse laser synchronously superimposed to the modulated cw pump beam. Pulses can be spatially aligned to either cavity, while the position of the cw pump beam is kept fixed at the center of the PM. Fig. \ref{switch_exp}a shows the initial (non-perturbed, top) situation.
As the short pulse is aligned at the left cavity, two events are observed: either L cavity was off before the arrival of the short pulse, i.e. the L0R1 state anticipated the pulse perturbation, in such a case L0R1 survives (Fig. \ref{switch_exp}a, center, dashed lines); or L1R0 state anticipated the pulse perturbation, in such a case a switch to L0R1 is observed (Fig. \ref{switch_exp}a, center, continuous lines). When the perturbation laser is shifted to the right cavity the situation is reversed: L0R1 switches to L1R0, while L1R0 remains unchanged. Note that a cavity in the on-state can be switched off with an extra pulse on it, which simultaneously switches on the adjacent cavity.
This pretty much resembles a light rocker switch in a house: "rocking" the lever by pushing it on the raised half makes the mechanism switching.
Pulsed external control is well reproduced by our theoretical model.

The short pulse laser can also be used to stabilize spontaneous switching such that photon trapping in one of the two cavities can be measured with a slower 2D detector. Fig. \ref{switch_exp}b shows intensity images captured on an InGaAs camera: the short pulse laser is used to stabilize either the L0R1 state with a pulse on the L cavity, or the L1R0 state with a pulse on the R cavity (see Fig. \ref{switch_exp}b, insets). These results further illustrate photon confinement states around the cavity regions. Switching asymmetric states is a clear advantage in the context of applications, for instance all-optical flip-flops in photonic integrated circuits\cite{Liu:2010fk}. This demonstration of coexistence of broken parity states through optical switching constitute an experimental proof of spontaneous mirror-symmetry breaking in a PM nanolaser.
A major interest of our system is that the photonic barrier amplitude and sign can be controlled by design\cite{Haddadi:14}. It can be shown that i) changing the sign of interactions is equivalent to reversing the coupling constant, an alternative to the control of the local nonlinearity through an external magnetic field by the Feshbach-resonance effect in BECs; ii) the symmetry of the bifurcated ground state can be exchanged without modifying the nature of interactions. In our case, the latter would be observable as far as the bonding state becomes the lasing mode.


In our experiments, the inferred photon number in each cavity is ${\cal S}\sim 100$ (the normalization photon number is ${\cal S}_{norm}\approx 135$) at the onset of SSB, compatible with the signal level measured by the ADP detectors (see Methods). Shrinking the middle row hole radius $r_{green}$ by just a few percent would result in a reduced mode splitting, from $K\tau= 3.3$ to, e.g., $K\tau \sim 0.7$. The predicted bifurcation point would then decrease to $\Delta P_c=K\tau/\alpha \sim 0.1$, which is 3\% far from laser threshold. This means that the pitchfork bifurcation point would be further shifted towards the laser threshold, eventually occurring on the steep portion of the S-curve. In such conditions, photon number in each cavity, ${\cal S} \approx {\cal S}_{norm} \Delta P_c$ becomes ${\cal S} \sim 10$ at the SSB instability. Quantum interference in PMs in presence of --even modest-- nolinearities are expected to leave its fingerprints on the quantum correlations of the laser photons\cite{Liew:2010uq}. This PhC molecule, combined with quantum dot technology\cite{faraon2008coherent}, might then constitute a building block for a news class of light emitting nano-sources with strong photonic correlations.

\textbf{\newline{}Methods}

\medskip{}

{\bf Sample fabrication}

The active membrane is grown by metalorganic chemical vapor deposition on an InP substrate with an intermediate InGaAs etch-stop layer and a SiO$_2$ sacrificial layer on top. It includes four InGaAs/InGaAsP quantum wells (photoluminescence centered at $\sim1510$ nm, FWHM =63 nm). This structure is bonded upside-down to a Si substrate coated with benzocyclobuten (BCB). The InP and etch-top layer are removed chemically leaving a structure composed of, from bottom to top: a $\sim 280\,\mu$m-thick Si substrate, a $\sim400$ nm-thick BCB layer, a $\sim1\, \mu$m-thick SiO2 layer and the 265 nm-thick membrane. Sample fabrication is achieved by deposition of a $\sim 200$ nm SiN layer (hard mask), e-beam lithography (2 nm-resolution) to write the PhC on a poly-methyl-methacrylate resist, and inductively-coupled-plasma reactive ion etching to etch the mask and the membrane. SiO$_2$ layer is removed chemically by AF acid penetrating the holes.

{\bf Setup description}

The nanolasers are pumped at $\lambda=808$ nm with a cw single-mode fibered laser diode (Lumics L808M100), modulated using a 120 MHz (AGILENT 81150A, minimum rise time $\sim2$ ns) waveform generator (pulses of few tens of ns, repetition rates from 10 to 200 kHz). A $\times100$ magnification, 0.95 numerical aperture (N.A) and IR antireflection coated microscope objective (OLYMPUS MPLAN 100xIR) is used to focus the pump on the sample down to a $1.5\,\mu$m spot diameter. Its relative position with respect to the cavities is adjusted (5 nm resolution) thanks to a nano-positioning sample holder (Melles Griot APT 600 6-axis stage) with piezoactuators and feedback loops. The emitted signal, collected through the same objective, is separated into three paths: i)  temporal analysis (see details below), ii)  spectral analysis with a spectrometer (Princeton Instruments, Acton SP2500i) coupled to a Ni cooled InGaAs 1D array detector (Princeton Instruments, OMA V, $\sim0.1$ nm resolution) and iii) IR imaging with an InGaAs camera (Sensors Unlimited SU 320) measuring both intensity and far-field emission profiles. The latter is obtained using a Fourier imaging technique (an additional lens on a kinematic base images the back focal plane of the objective).

{\bf Detection of laser emission}

The emission of each laser is measured separately using diffraction-limited spatial filtering. The beam in the temporal detection path is separated via a 50/50 beam splitter. Each beam is coupled into a single mode, $8\,\mu$m-diameter core, fiber through a $\times20$ (N.A. = 0.3, NACHET, IR coated) microscope objective. The fiber acts as a pinhole to select a specific region of the PhC. The ends of the fibers are connected to two identical 660 MHz-bandwidth, low noise avalanche photodiodes (APD, Princeton Lightwave PLA-841-FIB, equivalent noise power $\sim200\,\mathrm{fW}/\sqrt{Hz}$) to perform time domain measurements. The detection area is selected  by superimposing on the camera a 1550 nm laser spot, injected from the detection fiber, to a given nanolaser emission area. The cross-talk between the two detection channels (Ch1 and Ch2) has been quantified as follows: i) detection of the coupled nanolasers is optimized in Ch1 and Ch2; ii) sample is translated in the plane such that a single nanolaser of identical parameters is brought into one  detection area and maximized in Ch1; iii) a residual signal in Ch2 reveals a cross-talk of less than 10\% . The APD responsivity ($\sim60$ kV/W) and the signal level at bifurcation (7 mV) yield an optical power of 115 nW. Taking into account the transmission of optical elements and the coupling efficiency into the fiber, the output power emitted by one cavity (to a half space) is $P_{out}\approx1\,\mu$W. The equivalent total photon number inside the cavity is $S = 2\times P_{out}\tau/h\nu\approx110$ at the pitchfork bifurcation.

{\bf Peak detection algorithm}

Time traces in each channel are simultaneously recorded using a 13 GHz-bandwidth oscilloscope (Lecroy Wavemaster 813Zi) in the form of a sequence of 100 consecutive, 50 ns-duration time windows (one output pulse per window); 100 pulses are then recorded for each cavity in one shot. Within SSB conditions, two types of events are observed: a high pulse in cavity L (Ch 1) together with a low pulse in cavity R (Ch 2), called L1R0, and vice versa (L0R1). Such states appear either in long clusters (up to hundreds of pulses), or in segments of rapidly alternating L1R0-L0R1 events (few tens of pulses). We attribute long clusters to a small long-lived drift (typically due to mechanical vibrations), and segments with alternating events to spontaneous switching. The 600 $\mu$s-duration segment picked up in Fig. \ref{time_series}a is a typical example of spontaneous switching,  containing 31 pulses (50\% L1R0 and 50\% L0R1). A peak detection algorithm with threshold (75\% of the peak amplitude) is implemented in Ch2 to discriminate two cases: peak in Ch2 is larger than that in Ch1 (case L0R1), or smaller (case L1R0). Averages are performed over each type of events (Fig. \ref{time_series}b). When applying a short ($\sim\,100$ ps) perturbation pulse for demonstration of coexistence, peak detection is restricted to a time window starting with the pump pulse and ending at the occurrence of the perturbation pulse. Averages are subsequently performed (Fig. \ref{switch_exp}).


\bibliographystyle{apsrev4-1}

\begin{thebibliography}{30}
\bibitem [{gro()}]{ground}%
  \BibitemOpen
  \href@noop {} {}\bibinfo {howpublished} {The anti-bonding nature of the
  ground state in this molecule can be explained from the Bloch mode-effective
  refractive index of the underlying coupled W1 waveguide system (two adjacent
  missing rows of holes in the $\Gamma$K direction): the effective refractive
  index of the anti-symmetric mode is larger compared to the one of the
  symmetric mode, which is opposite to the usual situation in standard
  geometries (e.g. coupled ridge waveguides) \cite{Haddadi:14}}\BibitemShut
  {NoStop}%
\bibitem [{det()}]{detuning}%
  \BibitemOpen
  \href@noop {} {}\bibinfo {howpublished} {Explicit symmetry breaking due to
  frequency detuning ($\Delta \omega$) can be neglected in this PM since it is
  smaller than experimental error ($\Delta \omega/2\pi< 12$ GHz).}\BibitemShut
  {Stop}%
\end{thebibliography}
%


\textbf{\newline{}ACKNOWLEDGEMENTS}

We thank A. Amo, J. Bloch, S. Barbay, J. Dudley and A. Aspect for enlightening comments. This work was supported by the CNRS, ANR (ANR-12-BS04-0011) and the RENATECH network.

\end{document}